\documentclass[twocolumn,letterpaper]{aastex631}

\newcommand{\vw}{v_{\rm w}}
\newcommand{\vs}{v_{\rm shear}}

\newcommand{\cs}{c_{\rm s}}
\newcommand{\rs}{r_{\rm s}}
\newcommand{\rH}{r_{\rm H}}
\newcommand{\rB}{r_{\rm B}}
\newcommand{\Ro}{R_{\rm o}}
\newcommand{\RoH}{R_{\rm o,H}}
\newcommand{\Rp}{R_{\rm p}}
\newcommand{\Mp}{M_{\rm p}}
\newcommand{\RJ}{R_{\rm J}}
\newcommand{\lp}{\lambda_{\rm p}}
\newcommand{\lH}{\lambda_{\rm H}}
\newcommand{\xilos}{\xi_{\rm los}}

\usepackage{amsmath}

\begin{document}

\title{Streams and Bubbles: Tidal Shaping of Planetary Outflows}

\author[0000-0002-1417-8024]{Morgan MacLeod}
\affiliation{Institute for Theory \& Computation, Center for Astrophysics, Harvard \& Smithsonian, Cambridge, MA, 02138, USA}

\author[0000-0002-9584-6476]{Antonija Oklop{\v{c}}i{\'c}}
\affil{Anton Pannekoek Institute for Astronomy, University of Amsterdam, Science Park 904, NL-1098 XH Amsterdam, The Netherlands}

\author[0009-0008-4762-6170]{Fabienne Nail}
\affil{Anton Pannekoek Institute for Astronomy, University of Amsterdam, Science Park 904, NL-1098 XH Amsterdam, The Netherlands}

\author[0000-0001-6025-6663]{Dion Linssen}
\affil{Anton Pannekoek Institute for Astronomy, University of Amsterdam, Science Park 904, NL-1098 XH Amsterdam, The Netherlands}

\begin{abstract}
Planets lose mass to atmospheric outflows, and this mass loss is thought to be central in shaping the bimodal population of gaseous giant and rocky terrestrial exoplanets in close orbits. We model the escape of planetary atmospheres in three dimensional gas dynamic simulations in order to study their emergent morphology. Planetary outflows show a range of shapes from fast, isotropic outflows bounded by bow shocks to slower motion confined to thin streams. We show that a crucial factor is the role of the tidal gravity and orbiting reference frame in which planets lose mass. Flows can be characterized by the dimensionless Rossby number evaluated at the scale of the Hill sphere. Flows with a low Rossby number are significantly deviated and shaped by the stellar gravity, while those with a high Rossby number are comparatively unaffected.  Rossby number alone is sufficient to predict outflow morphology as well as kinematic gradients across transit. The known exoplanet population should span a range of outflow Rossby numbers and thus shapes. We can use this information to constrain outflow physics and to inform observing strategies. 
\end{abstract}


\section{Introduction}

Among the thousands of known exoplanets, a rapidly growing population have been identified with outflowing atmospheres \citep{2024A&A...689A.179O}. These systems are typically observed through their excess absorption during transit of their host stars, in either metal, hydrogen (eg. Lyman $\alpha$) or Helium (eg. the metastable helium line at 1083nm) lines. 
We can understand that these observations represent outflowing atmospheres (and model the inferred mass loss rates) when the excess absorption extends far above the planetary radius and has kinematics representative of escape from the planet's gravity \citep[e.g.][]{2004A&A...418L...1L,2009ApJ...693...23M,2012MNRAS.425.2931O,2018ApJ...855L..11O,2023MNRAS.518.4357O}.

At the simplest level, the observational strategy to identify excess absorption during transits has been to remove an out-of-transit spectral line profile from its in-transit counterpart. The implicit assumption here is that most of the excess atmosphere lies near the planet, so that when the planet is in transit, its escaping atmosphere will be as well. While this seems a priori reasonable, we are now learning that escaping planetary material can extend in huge streams ahead and behind an orbiting planet \citep[e.g.][]{2023SciA....9F8736Z,2024AJ....167..142G}.  The tens to hundreds of planetary radii extent of the excess absorption in these systems cannot be replicated with a spherical outflow. In particular, the out-of-transit absorption levels are far too high relative to mid-transit to be explained by a spherical ``wind" from the planet with $\rho\propto r^{-2}$ density fall off. Instead, a confined, geometrically-thinner but optically thicker stream-like configuration is needed to match the observed light curves of excess absorption versus orbital phase \citep{2024arXiv241019381N}.

This paper discusses the tidal shaping of planetary outflows, extending initial discussion by \citet{2010ApJ...721..923L} and \citet{2015A&A...578A...6M} and numerical work on the subject by \citet{2019ApJ...873...89M}. We emphasize that outflows are fundamentally a product of star--planet interaction and examine outflow dynamics in the star--planet combined gravity. We use hydrodynamic simulations of outflows to discuss how outflows form a variety of shapes, from extended streams to fast-moving bubbles bounded by shocks. In Section \ref{sec:scales}, we discuss characteristic scales in the star-planet gravity. In Sections \ref{sec:method} and \ref{sec:dynamics} we introduce our hydrodynamic modeling method and results. In Section \ref{sec:discussion} we discuss the possible implications and lessons that can be derived from planetary outflow morphologies, and in Section \ref{sec:conclusions} we conclude.

\section{Characteristic Scales}\label{sec:scales}

\subsection{A Planet in Isolation}

A planet in isolation launches a hydrodynamic wind when the thermal energy per unit mass of its outermost layers become comparable to the depth of the planet's gravitational potential well,  $G \Mp/\Rp$, where $\Mp$ and $\Rp$ are the planetary mass and radius. The thermal energy per unit mass is on the order of $\sim c_s^2$, where $c_s$ is the gas sound speed. The ratio of these quantities defines the planetary hydrodynamic escape parameter, 
\begin{equation}
    \lp = \frac{G\Mp}{\Rp \cs^2},
    \label{eq:lambdaP}
\end{equation}
where $\lp \gg 1$ implies a comparatively deep potential well and that the gas is strongly confined \citep[e.g.][]{1960ApJ...132..821P,1999isw..book.....L,2004A&A...418L...1L}. At smaller $\lp$, the thermal energy becomes larger relative to the depth of the potential, and a wind with speed $\vw \sim \cs$ can emerge. 

The structure of a such an outflow is subsonic near the planet and supersonic at large radii. If the sound speed is constant throughout the wind (which occurs if the flow is isothermal), then the sonic radius (where $\vw = \cs$) is located at 
\begin{equation}
    \rs \approx \frac{G\Mp}{2 \cs^2}= {\lambda_{\rm p} \over 2} \Rp  .  
\end{equation}
Thus, lower $\lp$ implies steeper pressure gradients that more quickly accelerate winds to the transonic point.

\subsection{Star--Planet Potential}

Crucially, planetary outflows exist in the mingled star--planet potential \citep[e.g.][]{2007A&A...472..329E,2019ApJ...873...89M,2023MNRAS.518.4357O}. The gravity of both components is important, and the planetary outflow emerges in an orbiting (and therefore accelerating frame). The planet is dominant over the gravitational potential within the Hill sphere (or Roche lobe) of approximate radius
\begin{equation}
    \rH = \left( \frac{M_{\rm p}}{3M_*} \right)^{1/3} a,
\end{equation}
which defines a depth of a potential, $GM_{\rm p}/\rH$, or a characteristic velocity, \begin{equation}
    v_{\rm H} \approx \left(\frac{GM_p}{\rH} \right)^{1/2} \approx \left(  \frac{3^{1/3} G M_p^{2/3} M_*^{1/3}}{a} \right)^{1/2}.
\end{equation}  
By analogy to $\lp$, we define the hydrodynamic escape parameter of the Hill sphere to be, 
\begin{equation}
    \lambda_{\rm H} = \frac{ GM_{\rm p}}{\cs^2 \rH }  = {v_{\rm H}^2 \over \cs^2} . 
\end{equation}
Where the second equality shows that the escape parameter also represents the relative size of the sonic radius as compared to the Hill sphere (additionally we note that $\lambda_{\rm H} = 2\rs /  r_{\rm H}$ for an isothermal outflow).

Additionally, the planet orbits with an angular speed $\Omega =  v_{\rm orb}/a = \sqrt{GM_*/a^3}$ (here and throughout, we  make the approximation that $M_* \gg \Mp$). As the planetary wind expands away from the planet in the rotating frame, it is diverted by the Coriolis acceleration, which imparts a vorticity $2\Omega$ into the rotating-frame flow.  For Keplerian orbits, the differential rotation $\Omega(r)$ implies a similar shear velocity that advects material forward or ahead of its original spatial location in the rotating frame. Over the length scale of $\rH$, the shear velocity is $\vs \sim \Omega r_{\rm H}  =  v_{\rm H}/\sqrt{3}$.  The dimensionless ratio of the Rossby number, $v / \Omega L$, where $v$ is a velocity and $L$ is a length scale, compares the importance of the wind flow from the planet to orbital advection due to the rotating reference frame in which the wind is lost from the planet. By adopting $\vw \sim \cs$, we define the Rossby number over the length scale of the Hill sphere as 
\begin{align}
    \RoH = \frac{\cs}{\Omega r_H} &= 3^{1/2} \frac{\cs}{v_{\rm H}} =  \sqrt{\frac{3}{\lambda_{\rm H}}}.
    \label{eq:rossby} 
\end{align}
The smaller the value of the Rossby number, the more significant the effect of orbital advection on the wind dynamics. Equation \ref{eq:rossby} shows that $\lH$ and $\RoH$ are not independent. Both are comparisons of the characteristic speeds of the Hill sphere and wind (via the approximation that $\vw \sim \cs$).

\subsection{Emergent Structures: Bubbles and Streams}

Considering the effects of both orbital shear and the depths of potential give us insight into the shaping of planetary outflows. \citet{2019ApJ...873...89M} offered insight into these dynamics by successively adding tidal and orbital terms to the equation of motion of a planet losing mass. 

When $\lambda_{\rm H} \ll 1$, the energy in the wind is much larger than the potential difference between the Hill sphere and infinity. This implies that the flow escapes the planet relatively uniformly, without regard to the Roche potential or locations of the Lagrange points. It also indicates a large Rossby number, $\Ro \gg 1$, such that the outflow will not be significantly affected by orbital shear on the scale of the Hill sphere. These cases should lead to relatively spherical ``bubble"-like outflows around a planet. 

However, when $\lambda_{\rm H} \gg 1$, the potential difference between the saddle points of the potential and other directions is significant relative to the wind energy, and escaping material is channeled through the inner and outer lagrange points. This indicates a small Rossby number $\Ro \ll 1$ and that orbital shear will dramatically deviate the outflow relative to that of an isolated planet. These cases lead to concentrated, thin ``streams" of planetary outflow.

\section{Numerical Method and Simulations}\label{sec:method}

\subsection{Hydrodynamic Method}

We use the Athena++ hydrodynamic code \citep{2020ApJS..249....4S} to perform simulations of the dynamics of gas outflowing from exoplanets. Our models parameterize the properties of the outflow by specifying planetary and stellar boundary conditions that are held constant through the duration of the calculation. These boundary conditions lead outflows into the active simulation mesh to develop and interact. 

Our methodology closely adheres to that described by \citet{2022ApJ...926..226M} and expanded upon by \citet{2024A&A...684A..20N,2024arXiv241019381N}, and our source code is available online at \cite{nail_2023_10025850}.

\subsection{Simulation Group}

The key model parameters in our calculations are the specification of the star and planet system and the escape parameter of the two outflows. We create a suite of calculations by varying these parameters. In each of our models we hold fixed the stellar mass, $M_\ast = M_\odot$, the planetary mass $\Mp = 10^{30}$~g~$ \approx 0.5 M_J$, the mass loss rates of the planet and star, both of which are set to $10^{11}$~g~s$^{-1}$. The stellar wind is further specified by the escape parameter $\lambda_{\ast} = 15$.    We then vary the orbital semi-major axis, $a$, planetary radius (in multiples of $R_J\approx 7\times 10^{9}$~cm), and the planetary hydrodynamic escape parameter, $\lp$. We adopt 3 values of each of these 3 parameters, giving 27 total parameter combinations, listed in Table \ref{tab:sims}.  

In each case, the numerical parameters are held fixed. In particular, the calculations' spatial descretization, which is described by a spherical polar mesh surrounding the star, with $(N_r,  N_\theta, N_\phi) = (192, 96, 192)$ zones. The radial extent of the mesh is from the stellar surface at $r=R_\odot$ to $1000R_\odot$, with logarithmically-spaced zones to preserve approximately-cubic zone shapes.
The angular extent is the full $4\pi$ of solid angle, from 0 to $\pi$ in $\theta$ and 0 to $2\pi$ in $\phi$.
We specify several regions of enhanced spatial resolution using  Athena++'s static mesh refinement capability. 
In a torus from $1.5\times 10^{11}$~cm to $5.75\times 10^{12}$~cm, we add one additional level (factor of two) of spatial resolution. We add additional levels of refinement (5 to 7 depending on semi-major axis) in a box extending $\pm 5R_J$ in every direction surrounding the planetary center. We specify nested boxes reducing the spatial refinement with distance from the planet until the torus refinement criterion is reached.  Together, these choices ensure sufficient resolution in the region where the wind develops near the planet (the minimum zone size around the planet is $\approx R_J / 16$ regardless of semi-major axis), and far from the planet where winds extend into the star--planet environment.

\begin{table}[tbp]
\begin{center}
\begin{tabular}{lcccccc}
a [au] & $\Rp$ [$R_J$] & $\lp$ & $\cs$ [km~s$^{-1}$] &  $\RoH$ & $\lH$ & $\xi_{\rm los}$ \\
\hline
0.025 & 0.5 & 2.0 & 30.6 & 2.94 & 0.35 & 0.48 \\
0.025 & 0.5 & 4.0 & 21.6 & 2.08 & 0.69 & 0.35 \\
0.025 & 0.5 & 8.0 & 15.3 & 1.47 & 1.39 & 0.04 \\
0.025 & 1.0 & 2.0 & 21.6 & 2.08 & 0.69 & 0.29 \\
0.025 & 1.0 & 4.0 & 15.3 & 1.47 & 1.39 & 0.02 \\
0.025 & 1.0 & 8.0 & 10.8 & 1.04 & 2.77 & -0.00 \\
0.025 & 2.0 & 2.0 & 15.3 & 1.47 & 1.39 & -0.05 \\
0.025 & 2.0 & 4.0 & 10.8 & 1.04 & 2.77 & -0.12 \\
0.025 & 2.0 & 8.0 & 7.6 & 0.74 & 5.55 & -0.37 \\
0.05 & 0.5 & 2.0 & 30.6 & 4.16 & 0.17 & 1.06 \\
0.05 & 0.5 & 4.0 & 21.6 & 2.94 & 0.35 & 0.82 \\
0.05 & 0.5 & 8.0 & 15.3 & 2.08 & 0.69 & 0.50 \\
0.05 & 1.0 & 2.0 & 21.6 & 2.94 & 0.35 & 0.79 \\
0.05 & 1.0 & 4.0 & 15.3 & 2.08 & 0.69 & 0.47 \\
0.05 & 1.0 & 8.0 & 10.8 & 1.47 & 1.39 & 0.15 \\
0.05 & 2.0 & 2.0 & 15.3 & 2.08 & 0.69 & 0.43 \\
0.05 & 2.0 & 4.0 & 10.8 & 1.47 & 1.39 & 0.16 \\
0.05 & 2.0 & 8.0 & 7.6 & 1.04 & 2.77 & -0.10 \\
0.1 & 0.5 & 2.0 & 30.6 & 5.88 & 0.09 & 1.00 \\
0.1 & 0.5 & 4.0 & 21.6 & 4.16 & 0.17 & 0.95 \\
0.1 & 0.5 & 8.0 & 15.3 & 2.94 & 0.35 & 0.96 \\
0.1 & 1.0 & 2.0 & 21.6 & 4.16 & 0.17 & 1.04 \\
0.1 & 1.0 & 4.0 & 15.3 & 2.94 & 0.35 & 1.00 \\
0.1 & 1.0 & 8.0 & 10.8 & 2.08 & 0.69 & 0.81 \\
0.1 & 2.0 & 2.0 & 15.3 & 2.94 & 0.35 & 0.95 \\
0.1 & 2.0 & 4.0 & 10.8 & 2.08 & 0.69 & 0.79 \\
0.1 & 2.0 & 8.0 & 7.6 & 1.47 & 1.39 & 0.46 \\
\hline
\end{tabular}
\end{center}
\caption{Parameters of hydrodynamic simulations, in which the semi-major axis, $a$, planetary radius $\Rp$, planetary escape parameter, $\lp$ are all varied. Other parameters, which are held constant, are discussed in the text. We also report the sound speed of the outflow $\cs$, the estimated Rossby number at the Hill sphere, $\RoH$, the Hill sphere escape parameter, $\lH$,  and the transit gradient of line-of-sight velocity, $\xilos$.   }
\label{tab:sims}
\end{table}

\subsection{Line-of-Sight Kinematics}

In general, the shape of a planetary outflow is not directly observable. Instead, our transit observations probe the structure of the gas as illuminated by the star, along a given line of sight. We therefore examine how the line-of-sight traces of bubble and stream morphologies differ.

To compute line-of-sight diagnostics in our spherical polar coordinate mesh (which is centered on the star), we sum along the radial direction of the mesh. The surface density is 
\begin{equation}
   \Sigma =  \sum \rho d r, 
\end{equation}
and the mass-weighted line-of-sight velocity is 
\begin{equation}\label{vlos}
   \bar v_{\rm los} =  \frac{1}{\Sigma}\sum -v_r \rho d r, 
\end{equation}
where $v_r$ is the gas radial velocity relative to the star. 

In what follows, we will use the total gradient of gas velocity integrated at the orientations of the observer at egress and ingress to transit, 
\begin{equation}
    \Delta \bar v_{\rm los} = \bar v_{\rm los,egress} - \bar v_{\rm los,ingress},
\end{equation}
and normalize this to the same gradient of planetary line-of-sight velocity at egress and ingress,
\begin{equation}
    \Delta  v_{\rm p,los} = \bar v_{\rm p,los,egress} - \bar v_{\rm p,los,ingress},
\end{equation}
defining the dimensionless line of sight velocity gradient to be 
\begin{equation}\label{xilos}
    \xi_{\rm los} = \frac{\Delta \bar v_{\rm los}}{ \Delta  v_{\rm p,los}}.
\end{equation}

\section{Outflow Dynamics}\label{sec:dynamics}

\subsection{Bubble and Stream Morphologies}

\begin{figure}
    \centering
    \includegraphics[width=0.5\textwidth]{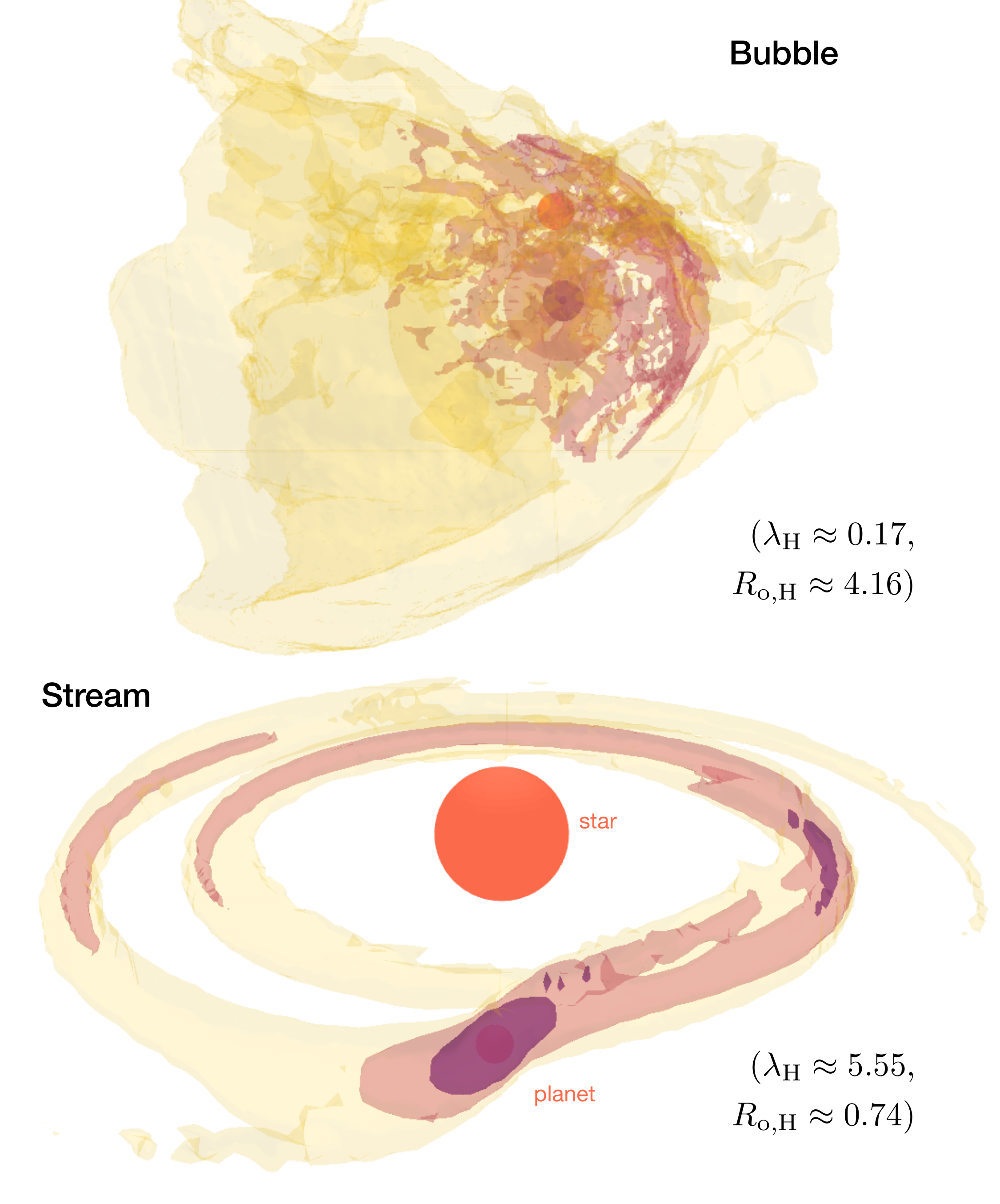}
    \caption{Volumetric renderings of two three-dimensional gas dynamic simulations of exoplanet outflows. Surfaces show constant densities of $\rho = 10^{-21}$, $10^{-20}$, and $10^{-19}$~g~cm$^{-3}$ in the upper, ``bubble" case, and $\rho = 10^{-17.5}$, $10^{-17}$, and $10^{-16.5}$~g~cm$^{-3}$ in the ``stream" case. The bubble model has a semi-major axis of $a=0.05$~au, a planetary radius of $\Rp=0.5\RJ$, and $\lp=2$. The stream model has $a=0.025$~au, $\Rp=2\RJ$, and $\lp=8$. In each example, the star and planet are marked with spheres, with the outflow originating from the planet. While in the bubble case, the outflow begins spherical, and is eventually shaped into a bow shock due to interaction with the stellar wind, in the stream case the outflow is quite directional, leaving from the inner and outer Lagrange points on the day and night side of the planet and stretching into long, thin streams that surround the star.  }
    \label{fig:3d}
\end{figure}

Figure \ref{fig:3d} highlights the difference in possible morphologies of planetary outflows. In the ``Bubble" case a $\lambda_{\rm p}=2$ flow emerges from a $0.5\RJ$ planet at $a=0.05$~au, implying a sound speed of $\cs \approx 31 $~km~s$^{-1}$, while in the ``Stream" case, a $\lambda_{\rm p}=8$ flow emerges from a $2\RJ$ planet at $a=0.025$~au, implying a sound speed of $\cs \approx 7.7 $~km~s$^{-1}$.  In both cases, the orbital velocity of the planet is considerably larger than these sound speeds, with $v_{\rm orb}\approx 133$ and $189$~km~s$^{-1}$, respectively.

The difference in the spatial configuration of the outflows is immediately apparent. The bubble is broadly distributed around the planet, forming a large-scale height outflow that extends in every direction radially away from the planet. While somewhat obscured by the turbulent limb in  Figure \ref{fig:3d}, a key feature of this flow is its symmetry about the planet.  The stream, by contrast, is comparatively geometrically thin but is denser, and thus optically thicker. It does not extend far from the orbital plane, but does stretch further along the orbital path than the bubble, both leading and trailing the planet. 

\subsection{Role of the Rossby Number}

\begin{figure*}
    \centering
    \includegraphics[width=\linewidth]{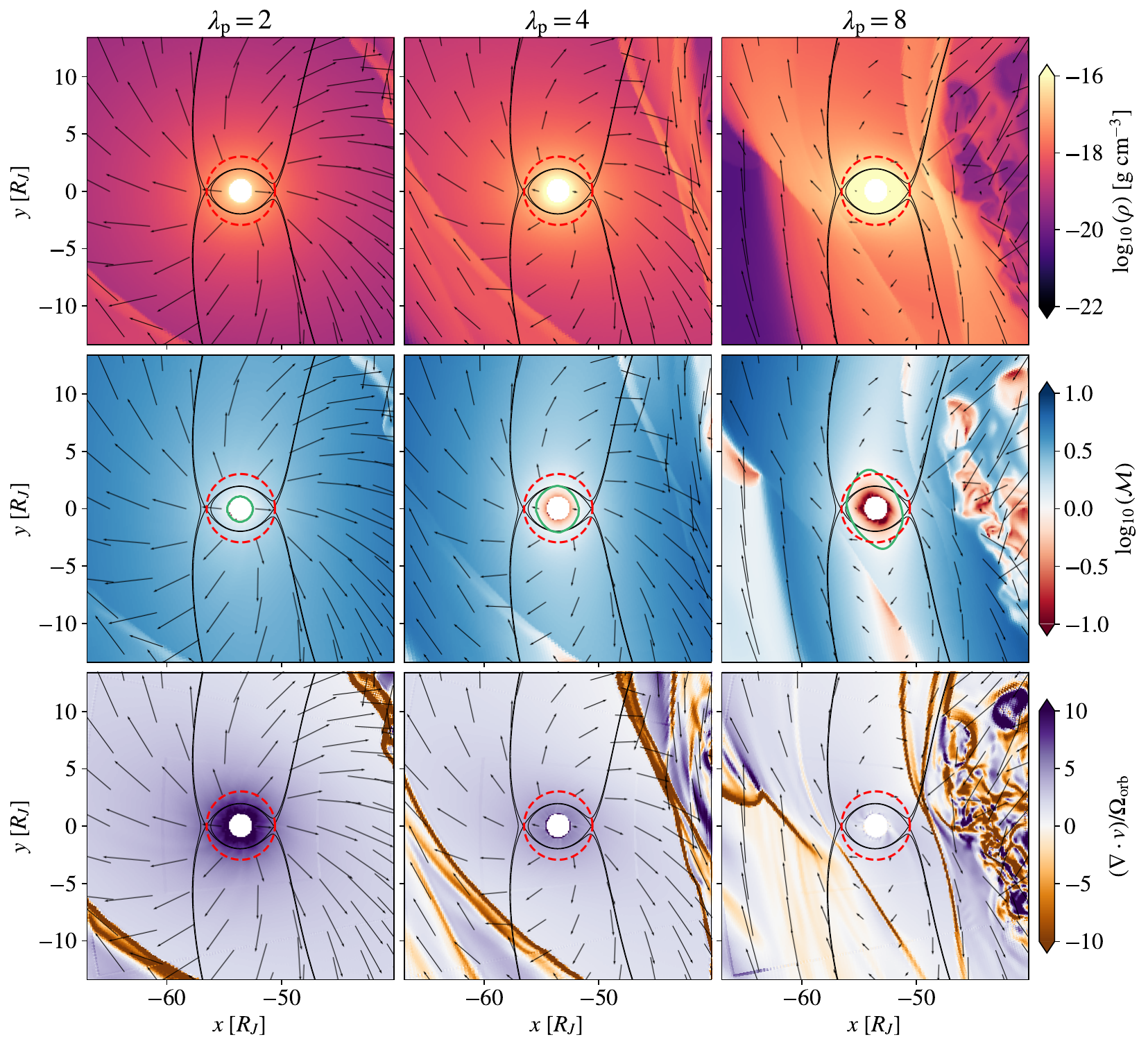}
    \caption{ Density, Mach number, and velocity divergence in slices through the orbital plane surrounding a planet losing mass. In each case, the planetary radius is $\Rp =\RJ$ and the orbital separation is $a=0.025$~au. The sound speed of the gas is the the only factor that changes, characterized by the hydrodynamic escape parameter $\lp$, from a hotter, faster outflow with $\lp=2$ to a cooler, slower with $\lp=8$. Arrows mark flow velocity in the corotating frame, black contours mark equipotential surfaces, the red dashed line is the Hill radius around the planet, and the green contour in the Mach number panel is the sonic surface.  Faster outflows (low $\lp$) are more spherically symmetric and radial relative to the planet, and less diverted by the curvature of the rotating frame. When velocity vectors are diverted to have similar perpendicular and radial magnitudes we see shocks emerge that bound a ``bubble" of free wind expansion around the planet.  }
    \label{fig:lp}
\end{figure*}

Between the extremes of fast, bubble-like outflows and slow-moving, extended streams there is a continuum of planetary outflow morphologies. Figure \ref{fig:lp} examines one such continuum, in which we show slices of the simulations in the $x-y$ plane of the orbit. We fix the system parameters to $\Rp = \RJ$, and $a=0.025$~au, and vary $\lp$ and thus the temperature and sound-speed of the planetary outflow. The panels of Figure \ref{fig:lp} show the density, Mach number, and velocity divergence of the outflow. 
The progression of $\lp=2,4,8$, corresponds to $\RoH \approx 2.08, 1.47, 1.04$. 
Velocities relative to the rotating and orbiting planetary frame are shown with vectors in Figure \ref{fig:lp}. Black contours trace the equipotential surfaces of the inner and outer Lagrange points and the red dashed circle marks the Hill sphere of the planet.  In the Mach number panel, the sonic surface, defined by Mach number of one, is shown with a green contour. 

First examining the case of $\lp=2$, we see that gas flows nearly radially away from the planet, slowly curving counterclockwise in the orbital plane due to the vorticity imparted by the rotating frame that the wind is is launched in. The planetary boundary condition is imposed at rest, but because of its high pressure, the outflow quickly accelerates to become supersonic, ${\cal M} > 1$, quite close to the planet, the sonic radius in the simulation is $\sim \Rp$, much like the predicted sonic radius for $\lp =2$, which  is $\rs \sim (\lp/2) \Rp = \Rp$. Thus, the acceleration of material to the transonic point happens well within the planet's Hill sphere -- where it's own gravity is more significant than the stellar gravity, so that the outflow is already moving supersonically when it crosses into the larger, stellar potential well. 

In the third row of Figure \ref{fig:lp}, regions of positive divergence indicate where the wind is expanding away from the planet.
The nearly-spherical outflow from the planet has a finite extent. At the border of this ``bubble," the velocity vectors are beginning to be significantly deviated from radial outflow relative to the planet by the Coriolis acceleration and orbital shear.  
Because the outflow is supersonic, a shock interface marks this transition, as highlighted by regions of strong flow convergence, $\nabla\cdot v \ll 0$ in Figure \ref{fig:lp}. The characteristic length scale of this re-direction of velocity vectors is the Rossby ``turning length," which implies bubble sizes
\begin{equation}\label{eq:rB}
    \rB \sim  \frac{\vw (\rB)}{ \Omega }, 
\end{equation}
where the notation $\vw (\rB)$ implies the wind velocity at the size scale of the bubble. In general, $\vw(\rB) \sim \cs$, but may be different by a meaningful factor of a few depending on how much larger or smaller the bubble is than the sonic radius in the accelerating outflow. Equation (\ref{eq:rB}) implies $\rB/a = \vw(\rB) / v_{\rm orb} \sim \cs / v_{\rm orb}$.  
As can be traced in velocity vectors, the combined effects of Coriolis acceleration and orbital advection dictates that bubbles are not spherical -- and extend further in the leading and trailing directions than toward or away from the star. 

As $\lp$ increases, and the outflow becomes cooler and slower, the morphology of mass loss changes.  Rather than an isotropic, spherical outflow, the material is increasingly channeled into two tidal tails -- one lost primarily from the day side of the planet near the $L_1$ Lagrange point and one lost from the night side of the planet near the $L_2$ Lagrange point. Especially for $\lp =8$, the orbiting frame in which gas is lost from the planet is   significant in shaping the flow. By the time the outflow expands to the scale of the Hill sphere, it is already significantly deviated from a radial outflow to stream ahead and behind the planet along the orbital path. We can understand this result in terms of the Hill sphere Rossby number, $\RoH$, which decreases as $\lp$ increases. 

For $\lp=8$, $\RoH\sim 1$, which means the effect of the orbiting frame is similar to that of the winds own expansion.  This means that before the wind can expand significantly from the planet it is deviated by the coriolis acceleration and then sheared away into tidal tails by the differential rotation of Keplerian orbits, $\Omega(r)$. An additional point to notice is the differing dynamics of the streams lost near the inner (day side) and outer (night side) Lagrange points. The inner stream expands to smaller separations from the host star, where the orbital angular velocity, $\Omega$ increases. This means that material lost in this direction is sheared forward in the corotating frame -- moving ahead of the planet roughly along the orbital path. Material lost from the night side moves with slower angular frequency than the planet and trails behind. In a situation where the day and night sides of the planet are otherwise equal, this creates dual streams extending in both the leading and trailing directions. 

In the center panels for $\lp=8$, we see that the sonic surface is shaped by the stellar potential. Indeed, in this case, the predicted $\rs > \rH$, but gas begins to be accelerated by the stellar gravity rather than its own pressure gradient and the sonic crossing lies around the Hill sphere or Lagrange points -- as is found in the dynamics of streams of gas flowing through Lagrange points in binary star mass transfer \citep{1975ApJ...198..383L,2010ApJ...721..923L}. 

In the limiting case of very slowly-expanding, thin streams (e.g. as shown in Figure \ref{fig:3d}), the characteristic scale of spreading along the streams is related to the criterion of equation (\ref{eq:rB}). But because the stream doesn't have the chance to accelerate to $\vw > \cs$ within its own Hill sphere, we find 
\begin{equation}\label{eq:Hstream}
    H_{\rm stream} \sim \frac{\cs}{\Omega},  
\end{equation}
where $H_{\rm stream}$ represents the typical vertical and radial extent of the column-like stream. Along the orbital path, the stream is sheared so that its length is $\gg H_{\rm stream}$.

\subsection{Outflows in Phase Space}\label{sec:phasespace}

\begin{figure}
    \centering
    \includegraphics[width=\linewidth]{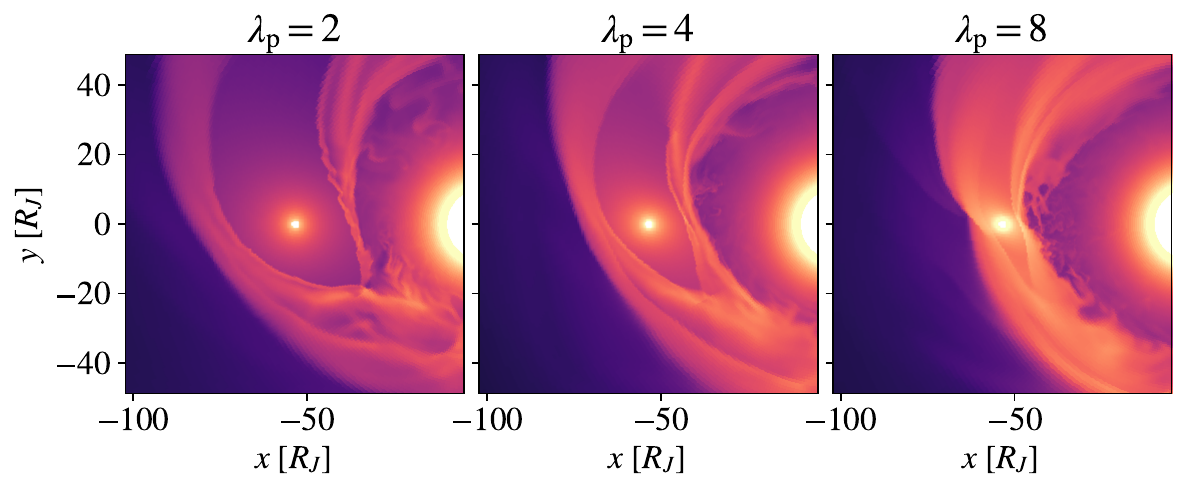}
    \includegraphics[width=\linewidth]{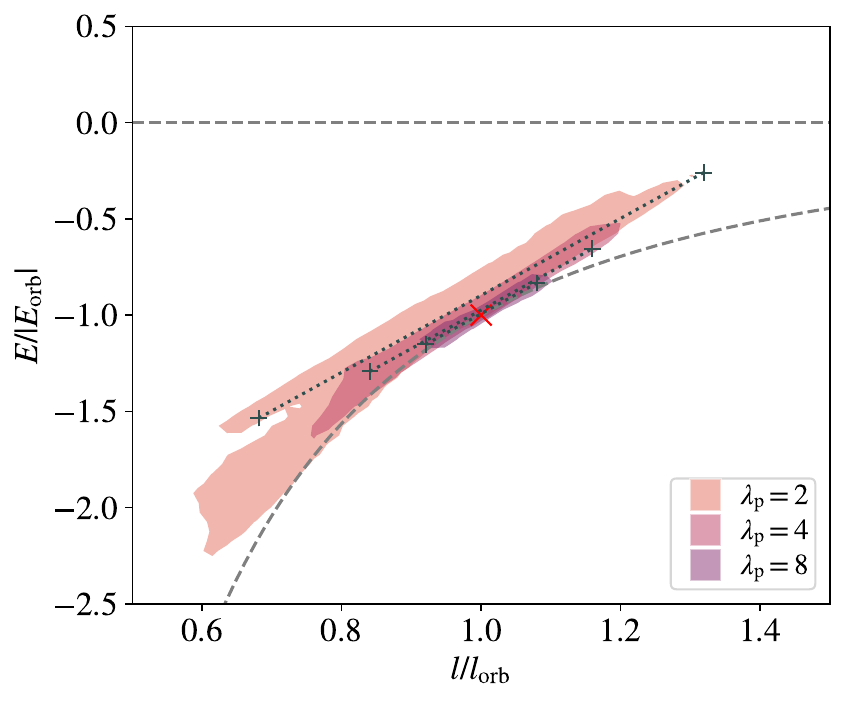}
    \caption{The model outflows of Figure \ref{fig:lp} in energy--angular momentum phase space. The upper panels show a broader view of the density structure around the planet, and the lower panel shows contours of the distribution of gas in energy and angular momentum space, normalized to the location of the planet (marked with a red X). Energy of zero is marked with a dashed line, as is the locus of circular orbits. Hotter, faster outflows have broader kinetic energy and  angular momentum spread (lower $\lp$), while cooler, slower outflows (higher $\lp$) cluster closely around the phase-space location of the planet, despite their spatial stretching into long streams. Finally, the outflows organize along discrete lines in energy-angular momentum space because of the superposition of the outflow direction and orbital motion. The dotted lines mark this spread, described by equations \eqref{dE} and \eqref{dl}, for wind velocities of 15, 30, and 60~km~s$^{-1}$ (which have the narrowest to broadest spread, respectively).    }
    \label{fig:EJ}
\end{figure}

We opened our discussion of characteristic scales for planetary outflows (Section 2) by comparing characteristic potential wells -- that of the planet and the Hill sphere -- to the energy of the planetary outflow. Indeed we find that considering the properties of planetary outflows in energy-angular momentum phase space paints a richer picture of the origins for their dynamics. 

 The specific energy of a fluid parcel is 
\begin{equation}
    E = \frac{v_i^2}{2} - \frac{GM_*}{r_*} - \frac{G\Mp}{r_{\rm p}}, 
\end{equation}
where $v_i^2/2$ is the kinetic energy in the inertial (not the corotating) frame, $-GM_*/r_*$ is the stellar potential based on the distance $r_*$ from the star, and $-G\Mp/r_{\rm p}$ is the planetary contribution to the potential given the distance from the planet, $r_{\rm p}$. The specific energy can be compared to 
\begin{equation}
    E_{\rm orb} = - \frac{GM_*}{2a}, 
\end{equation}
the specific orbital energy of the planet. 

The specific angular momentum of a fluid parcel is 
\begin{equation}
    l = \vec{r}_* \times \vec{v}_i ,
\end{equation}
where velocities $\vec{v}_i$ are again evaluated in the inertial frame rather than the corotating frame of the simulation. Specific angular momenta can be compared to 
\begin{equation}
    l_{\rm orb} = \sqrt{G M_* a},
\end{equation}
the specific angular momentum of the orbit. 

A planetary outflow begins at the planet, and therefore has the specific angular momentum and energy of the planet. The distribution in $E-l$ phase space is broadened by two effects, the  spread of initial energies and angular momenta due to the finite radius of the planet, and the spreading that occurs as pressure gradients accelerate the planetary wind. When we assume $\Rp \ll a$ and that the wind velocity is similar to its sound speed and $\vw \ll v_{\rm orb}$, the initial spread in specific energy due to the planetary radius is 
\begin{equation}\label{dE}
    \frac{\Delta E}{E_{\rm orb}} \sim  \pm \left( \frac{\Rp}{a} + \frac{2 \vw}{v_{\rm orb}} \right),
\end{equation}
while the spread in specific angular momentum is 
\begin{equation}\label{dl}
    \frac{\Delta l}{l_{\rm orb}} \sim \pm \left( \frac{\Rp}{2a} + \frac{\vw}{v_{\rm orb}} \right), 
\end{equation}
where the first term in each expression represents the spread due to the finite planetary radius, and the second due to the launching of the planetary wind. 

If we associate $\vw \sim \cs$, then we can compute the estimated dispersions above for the ``bubble" and  ``stream"  examples of Figure \ref{fig:3d}. In the bubble case, $\Rp/a\sim 0.01$, while $\cs / v_{\rm orb} \sim 0.16$, so the wind's outflow velocity dramatically regulates the dispersion of  energy and angular momentum. By contrast, in the stream case, $\Rp/a \sim 0.04$ and $\cs / v_{\rm orb} \sim 0.04$, so the initial shape of the planet and the eventual wind contribute similarly to the dispersion in phase space. Secondly, the magnitudes of these quantities show that the hot outflow of the bubble case achieves a much larger phase space spread than the stream does. 

Figure \ref{fig:EJ} examines the phase space and orbital distributions of the models of Figure \ref{fig:lp}, with increasing $\lp$. When $\lp$ is low, $\cs$ is high. In this case, wind energies are large enough to traverse the local potentials easily and expand into the environment. They also represent a broad distribution in energy-angular momentum phase space. By contrast, when $\lp$ is high, $\cs$ is small. The planetary outflow is narrowly confined in energy and angular momentum space. This means that fluid trajectories are increasingly like collisionless, ballistic orbits. The morphology of the resulting streams is set by the spread in the planetary surface energy and angular momentum, and the eventual flow, though stretched and sheared in physical space, is narrow in phase space, sharing orbital elements with the planet that created it.

\subsection{Line-of-Sight Kinematics}

\begin{figure*}
    \centering
    \includegraphics[width=0.49\textwidth]{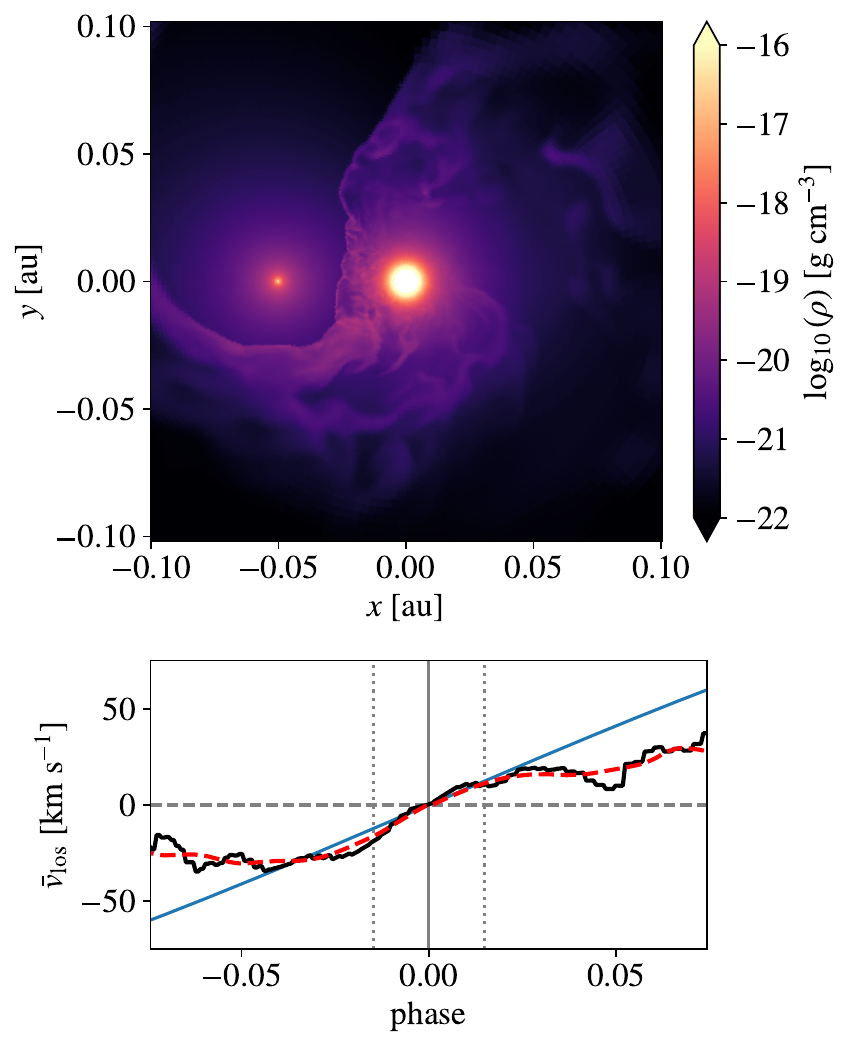}
    \includegraphics[width=0.49\textwidth]{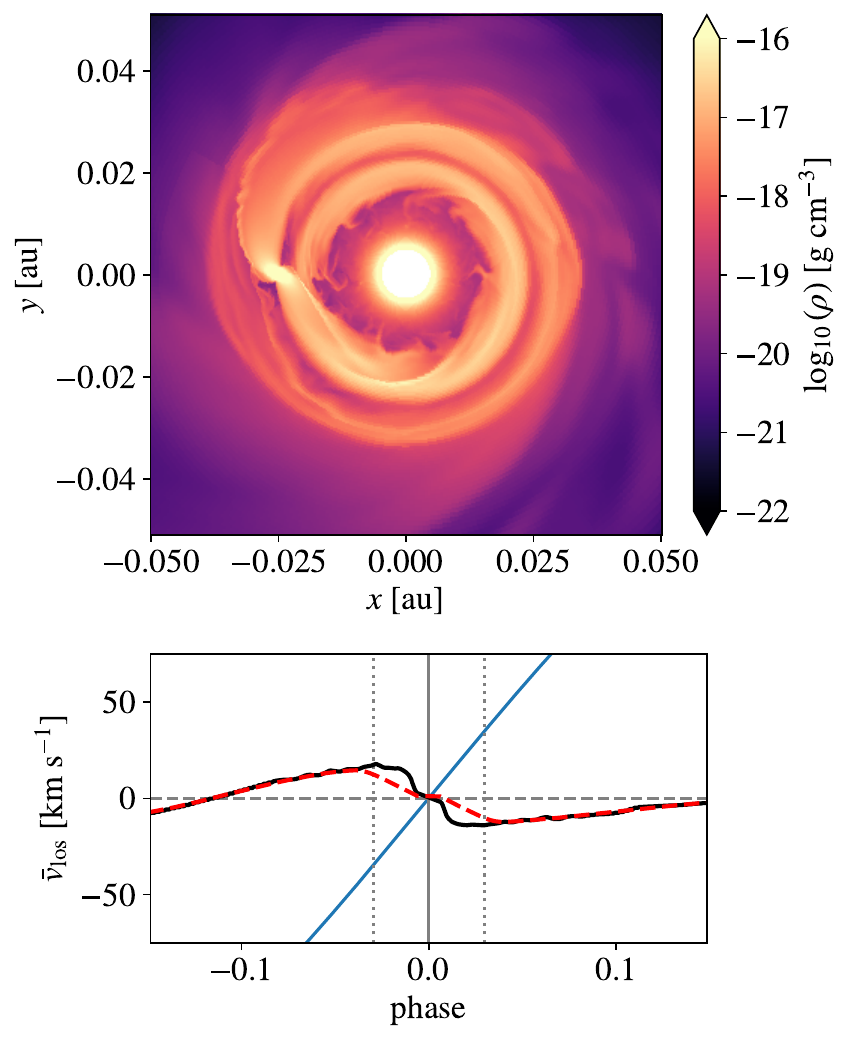}
    \caption{The bubble (left) and stream (right) outflows of Figure \ref{fig:3d} in slices of density through the orbital plane. Here we also show the projected line-of-sight velocity to an observer near the optical transit (phase of zero, with extent marked by vertical lines). The black line shows the mass-weighted line-of-sight velocity at a given angle and the red-dashed line averages over a range of angles that  account for a finite stellar disk. The lighter, blue line traces the planetary motion. The line of sight kinematics of bubble-like and stream-like outflows are quite distinct. Spherical bubble-like outflows emerge equally in all directions from the planet, and the averaged velocity tracks the planetary velocity closely, from moving toward the observer pre-transit (negative $\bar v_{\rm los}$) to moving away from the observer after transit (positive $\bar v_{\rm los}$). The stream outflow has opposite kinematics, with the leading stream moving toward the star and away from the observer, while the trailing stream moves away from the star and toward the observer.  }
    \label{fig:los}
\end{figure*}

We have so far described planetary outflow morphologies in terms of their three-dimensional shapes surrounding a host star. While these  morphologies are directly visible in a simulation, they are not directly observable in real systems. Instead, we turn to a one-dimensional tracer of outflow morphology, the line-of-sight kinematics of the outflow as a function of orbital phase, as defined by equation \eqref{vlos}. We emphasize, however, that a simple mass weighted line of sight velocity is not directly observable -- a real observation of a spectral line will have dependence on the line depth and the resulting line forming region in physical space around the planet \citep[e.g. as discussed by][]{2024arXiv241019381N}. But, our analysis is most applicable to the fully optically thin limit.  Because these complexities are quite system-specific, we leave their examination to future work on particular system case studies.

Figure \ref{fig:los} displaces density slices in the orbital plane of the ``bubble" and ``stream" calculations of Figure \ref{fig:3d}. In these two cases, the lower panels display $\bar v_{\rm los}$ as a function of phase. In these lower panels, the blue solid line shows the motion of the planet from moving toward the observer (negative line of sight velocity) to moving away from the observer (positive line of sight velocity). Vertical dotted lines mark the extent of the transit ingress to egress. Black lines measure $\bar v_{\rm los}$ as a function of phase, and the red dashed lines show the same $\bar v_{\rm los}$ but averaged over the phase-coverage of the stellar disk. 

What we observe in the line of sight velocity is that the bubble and stream morphologies are quite distinct in their slope with phase. At mid transit, both systems tend toward zero line of sight velocity. But they exhibit opposite gradients with phase. The bubble morphology more or less traces the motion of the planet. Material is lost isotropically around the planet, and so the averaged velocity just reflects the motion of the planet plus the dispersion of isotropic expansion. In the stream like case, the outflow is strongly shaped by the tidal gravity of the star. The leading stream, which transits the star at negative phases, is lost from the day side of the planet and falls through the inner Lagrange point between the planet and star, and exhibits motion toward the star (positive line of sight velocity). After mid transit, we are tracing material in the trailing stream, which is preferentially lost from the night side of the planet and which moves through the outer Lagrange point. This gas has negative line of sight velocities representing motion toward the observer and away from the star.

\begin{figure}
    \centering
    \includegraphics[width=\linewidth]{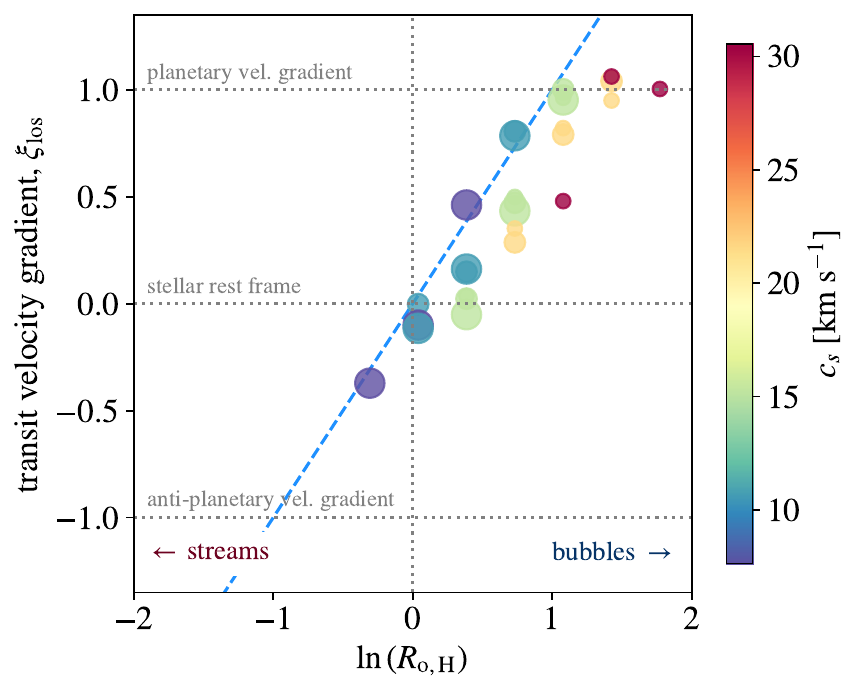}
    \caption{Dimensionless line-of-sight velocity gradient, $\xilos$, equation \eqref{xilos}, in terms of Hill-sphere Rossby number for all of our simulation set. Points are colored by sound speed and sized by planet radius. Despite widely-varying initial parameters, we observe that all of the models collapse in this parameter space, approximately following a simple relationship where Rossby number predicts the outflow morphology and kinematics as measured by $\xilos$. }
    \label{fig:xi}
\end{figure}

The difference in these line of sight velocity gradients allows us to use them as a simplified indicator of the overall outflow shape. Figure \ref{fig:xi} compares the dimensionless line of sight velocity gradient, $\xilos$, defined in equation \eqref{xilos}, to the hill sphere Rossby number, $\RoH$. We plot each of the simulated models, using color to denote outflow sound speed, and marker size to show planetary radius.  As $\xilos \rightarrow 1$, this means that the outflow line-of-sight velocity follows the planet's motion. $\xilos \sim 0$ implies no velocity gradient, and $\xilos < 0$ implies an inverted velocity gradient, where the motion of the outflow opposes that of the planet as a function of phase. 

In sum, the free parameters of our simulation group largely collapse when shown in these dimensionless ratios. The hill sphere Rossby number is a reasonable predictor of line-of-sight velocity gradient, and therefore, overall outflow shape around the planet. Our models lie close to the line $\xilos \sim \ln \RoH$, saturating at $\xilos \rightarrow 1$ for high Rossby number in the bubble-like limit. Models in the stream-like case have lower Rossby numbers and show much shallower or even inverted velocity gradients.

\section{Discussion}\label{sec:discussion}

\subsection{What can Outflow Shapes Teach Us?}

We have shown that planetary outflow morphologies sensitively probe the star--planet environment. They are reactive to the stellar wind ram pressure \citep{2022ApJ...926..226M}, and they are shaped by the tidal gravitational forces in which they are launched \citep{2019ApJ...873...89M}. In particular, we find that the Hill sphere Rossby number is a useful predictor of our simulated outflow morphologies. We further show that morphologies imprint strongly in outflow kinematics along the line of sight. Taken together, these conclusions suggest that it may be able to use measured constraints on outflow morphologies -- from light curve fitting or phase-resolved kinematics -- and convert these into constraints on the sound speed and thermodynamics of a planetary outflow. A too-hot outflow forms a bubble not a thin, confined stream. If observational evidence supports a particular morphology, this may provide significant meaningful constraints on the temperature of the planetary outflow which are distinct from a spectral FWHM, for example.

\subsection{Implications for Observing Strategies}

One implication of our findings is that the shapes of planetary outflows are quite varied, and, especially in stream-like cases there excess absorption can extend well beyond a planet's optical transit. The reasons can be seen particularly clearly in Figure \ref{fig:los}, where the density drops off quickly away from the planet in the bubble-like case, while in the stream case the column-like shape of the streams means that the outflow density does not drop off nearly as steeply with distance from the planet.  If excess absorption extends beyond optical transit, then we need to carefully consider the strategy for establishing the baseline absorption level of a particular spectral tracer -- a simple ``in" vs "out" of transit comparison may not suffice to characterize the full line depth if excess absorption contaminates the out of transit spectrum. Of course, the need for long baselines is in tension with the desire to observe in a brief window in order to mitigate intrinsic stellar variability. We suggest that it may be useful to consider a predicted morphology (described by the Hill sphere Rossby number) when planning an observing strategy for a given system. In the status quo, we are perhaps biasing our measurements of stream-like outflows by attributing part of the planetary signal to the stellar line.

\subsection{Predictions for the Exoplanet Population}

\begin{figure*}
    \centering
    \includegraphics[width=0.8\linewidth]{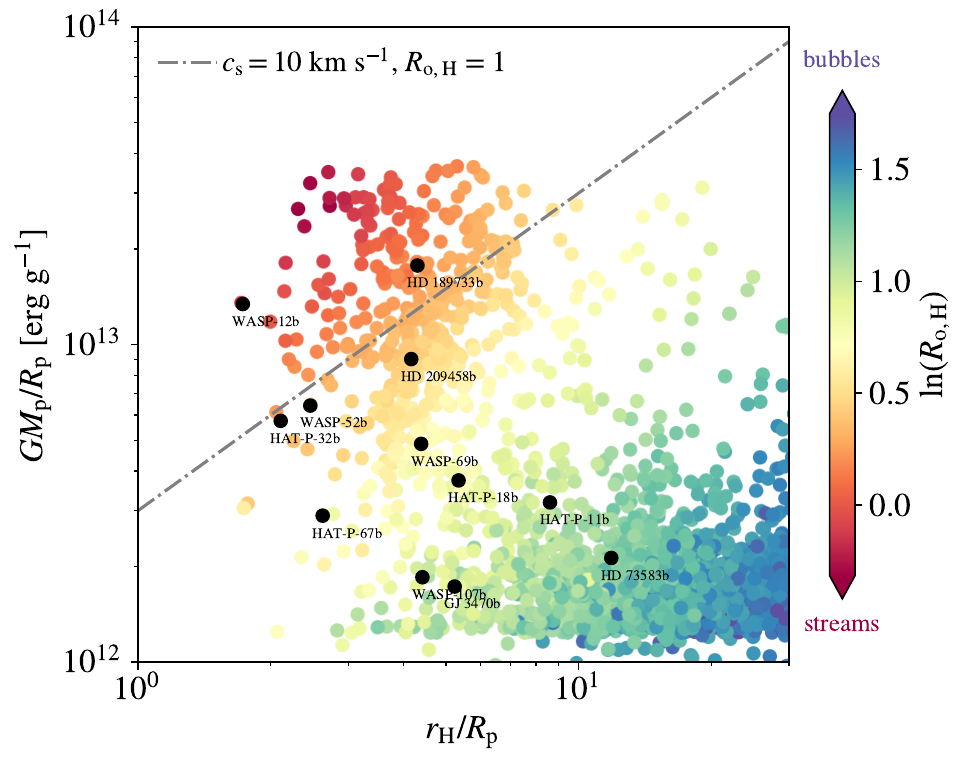}
    \caption{The gravitational potential of the planet versus the ratio of the Hill and planet radius. Colored points mark exoplanets with known mass and radius, from the {\tt sunset} catalog of predicted exoplanetary photoevaporative outflow properties, selecting planets with radius of greater than $1.6R_\oplus$ \citep{2024arXiv241003228L}.  Planets with detected signatures of atmospheric escape, either from Lyman-$\alpha$ or He 1083~nm line observations, are indicated by black circles. Colors indicate the predicted hill sphere Rossby number for known planets, where $\ln( \RoH) > 0$ predicts bubble-like morphologies and $\ln(\RoH) < 0$ predicts stream-like morphologies. The known exoplanets are widely distributed in predicted Rossby number, with examples in both extremes and a continuum between. The dashed grey line marks an outflow with a sound speed of 10~km~s$^{-1}$ and $\RoH=1$ for context. }
    \label{fig:population}
\end{figure*}

In Figure \ref{fig:population} we show systems from the NASA exoplanet archive with measured planetary mass, radius and stellar mass. We plot these systems in the space of Hill sphere size relative to the planet, $\rH/\Rp$, versus planetary potential, $G\Mp/\Rp$. For each planet, we draw on the {\tt sunset} catalog of predicted photoevaporative outflow properties to estimate a characteristic outflow temperature \citep{2024arXiv241003228L}. From the temperature, we estimate a sound speed $\cs^2 = \gamma kT/(\mu m_{\rm p})$, where $\gamma \approx 5/3$ is the adiabatic index, $k$ is the Boltzmann constant, $\mu$ is the mean molecular weight, and $m_{\rm p}$ is the proton mass.  Typical sound speeds range from 2.5 -- 15~km~s$^{-1}$. With the sound speed estimated, we then compute the predicted Hill sphere Rossby for each planet, and color points accordingly. In predicting outflow temperatures, the {\tt sunset} catalog needs to make assumptions about the unknown stellar spectral energy distribution and outflow metallicity (assumed to be solar), as well as other modeling choices with the {\tt sunbather} code \citep{2024A&A...688A..43L}. If the true temperatures are cooler than predicted by {\tt sunbather}, the Rossby number decreases, while if they are hotter, it increases, as described by equation \eqref{eq:rossby}. 

The space of Hill sphere filling fraction versus planetary potential maps out a dichotomy of predicted Rossby number. Planets with a deeper potential that fill large fractions of their Hill spheres (upper left) tend to have  Rossby numbers on the order of unity, or $\ln(\RoH) \sim 0$, because they are also the systems in the tightest orbits around their host stars. For these planets, we predict uniformly stream-like outflows, and that the tidal gravity will be very important.  By comparison, planets that fill small fractions of their Hill sphere ($\rH/\Rp\gg 1$), or those that have lower potentials, tend to have $\ln(\RoH) >0$ and we predict bubble-like outflows without much influence from the tidal gravity.  For context, a line marks where a planet with a sound speed of of 10~km~s$^{-1}$ and $\RoH=1$ but varying $\rH/\Rp$ would lie in this parameter space. This highlights that, as noticeable as the extremes are, there is a continuum of predicted Rossby number, with many systems with intermediate values lying approximately along this diagonal line.

Finally, we can compare to systems with detected features of atmospheric evaporation. HAT-P-67b and HAT-P-32b have both been argued to show strongly stream-like outflows \citep{2023SciA....9F8736Z,2024AJ....167..142G,2024arXiv241019381N}. In our parameter space, these lie in regions with predicted moderate (HAT-P-67b) and stream-like (HAT-P-32b) Hill Sphere Rossby number. \citet{2024arXiv241019381N} have recently used three-dimensional models to show that to achieve the geometry inferred from 1083~nm He absorbtion and kinematic curves as a function of orbital phase, HAT-P-67b and HAT-P-32b must have even cooler outflows than predicted from {\tt sunbather} (by a factor of $\lesssim 2$) effectively lowering their Rossby number and leading to a more-extended stream-like geometry. Additionally, the observations favor mass loss preferentially from the planetary day sides. By contrast, HD 73583b, shows evidence for excess absorption primarily during optical transit, as predicted by Figure \ref{fig:population}. HD 73583b has been previously modeled with an outflow closely confined around the planet, in part due to the ram pressure of the stellar wind \citep{2022AJ....163...67Z}. Thus, clearly there are subtleties (or at least particularities for individual systems), but the overall confirmation of predicted outflow geometry as a function of phase-space location is encouraging.   

\subsection{Shaping by Stellar Winds}

Finally, we revisit the crucial importance of shaping by stellar winds, as discussed by \citet{2022ApJ...926..226M}. When the stellar wind increases relative to the planetary wind, any of the initially stream or bubble-like flows are re-directed into confined, cometary tails. We find that the dynamics of these tails is mostly dictated by the ram pressure of the stellar wind, and is therefore similar to the description of \citet{2022ApJ...926..226M} regardless of the flow Rossby number at the Hill sphere. The only quantitative difference is the geometrically-thinner stream-like flows present a smaller cross section to stellar winds and are therefore somewhat more difficult to ablate and redirect \citep[For a concrete example, see Figure 6 of ][]{2024arXiv241019381N}. In phase space (e.g. as shown in Figure \ref{fig:EJ}, the impact of stellar winds sweeps planetary material from the confined tracks toward lower binding energy and lower angular momenta. Comparatively high planetary mass loss rates yield outflows unperturbed by the stellar wind's ram pressure, while lower planetary mass loss rates are more likely to be impinged upon.

\section{Conclusions}\label{sec:conclusions}

This paper has studied the role of stellar tidal gravity in shaping exoplanetary outflows. Coupled with an understanding of the effects of the stellar winds \citep[e.g.][]{2022ApJ...926..226M}, anisotropic mass loss \citep[e.g.][]{2009ApJ...694..205S,2024A&A...684A..20N,2024arXiv241019381N} and magnetic interactions \citep[e.g.][]{2009ApJ...704L..85C,2011ApJ...733...67C,2011ApJ...730...27A,2014MNRAS.444.3761O,2020ApJ...890...88O,2020ARep...64..259Z,2021MNRAS.508.6001C,2024MNRAS.527.5117S}, these effects shape a growing picture of planetary mass loss in an intertwined star--planet environment \citep{2015A&A...578A...6M}.  Observations of planetary mass loss inform how planetary populations are shaped by this effect \citep[e.g.][]{2013ApJ...775..105O,2017ApJ...847...29O,2018MNRAS.479.5012O}, but also may need to account for differences in outflow geometry between distinct systems. 
Our main findings are: 
\begin{enumerate}
    \item Planetary outflows can form a variety of shapes in the star--planet potential, from spherical ``bubble"-like outflows to extended, thin streams tracing along the orbital path (Figure \ref{fig:3d} and \ref{fig:lp})
    \item Stream like outflows are kinematically cool despite being spatially extended, in energy-angular momentum phase space they occupy a small spread around the planet's orbital energy and angular momentum. More spherical, bubble-like outflows are kinematically hotter and have broader phase-space dispersions (Figure \ref{fig:EJ}). 
    \item The key difference that shapes these outflows is the ratio of the outflow speed to the importance of the orbiting frame that outflows are launched in, we adopt the Hill sphere Rossby number, equation \eqref{eq:rossby} as a dimensionless metric.  The Hill sphere Rossby number also predicts outflow line-of-sight kinematics, like the velocity gradient across a transit (Figures \ref{fig:los} and \ref{fig:xi}). 
    \item The known population of exoplanets includes systems with a wide range of predicted Rossby numbers. There should be exoplanet outflows with a broad range of geometries, including among the systems detected to date (Figure \ref{fig:population}). 
\end{enumerate}

The promise of a deeper understanding of planetary outflow morphologies is linking these shapes to outflow thermodynamics and kinematics. At the root of these properties lie the important questions of outflow physics and launching mechanisms. If outflows are cool, they tend to have confined, stream-like three-dimensional shapes and line-of-sight kinematics. If they're hot, they're more spherical around the planet, rapidly dropping in surface density as the flow expands. Transit spectra probe excess absorption and its Doppler shifts as a function of orbital phase and we can hope to directly unravel these morphologies from observables. Doing so for a particular system or spectral line likely still requires catered radiative-transfer calculations because of the non-linearity of absorption as a function of optical depth. Nonetheless, doing so offers a profound constraint on planetary outflow sound speeds (and therefore temperatures) independent of assuming that spectral line widths amount to thermal broadening. Thus, we hope that the variety of planetary outflow morphologies, and their imprint in transit observations, can become useful tools in our effort to uncover the physics of planetary atmospheric mass loss.

\vspace{1cm}
We gratefully acknowledge helpful conversations with M. Holman, A. Loeb, R. Murray-Clay, J. Owen, J. Spake, S. Vissapragada. M.M. is grateful for support from a Clay Postdoctoral Fellowship at the Smithsonian Astrophysical Observatory. The simulation models in this work used stampede2 and stampede3 at the Texas Advanced Computing Center of the University of Texas through allocation PHY230031 from the Advanced Cyberinfrastructure Coordination Ecosystem: Services \& Support (ACCESS) program, which is supported by U.S. National Science Foundation grants \#2138259, \#2138286, \#2138307, \#2137603, and \#2138296.  A. Oklop\v{c}i\'{c} acknowledges support from the Dutch Research Council NWO Veni grant.

\software{  \texttt{astropy} \citep{astropy:2013, astropy:2018, astropy:2022}, \texttt{matplotlib} \citep{Hunter:2007}, \texttt{numpy} \citep{numpy}, \texttt{python} \citep{python}, \texttt{scipy} \citep{2020SciPy-NMeth, scipy_12522488}, and \texttt{Athena++} \citep{2020ApJS..249....4S}.
The Software Citation Station \citep{software-citation-station-paper}. 
The software to reproduce this work is available online at \url{https://github.com/morganemacleod/PWMorphology}, and will be archived with a DOI following peer review. 
}

\bibliographystyle{aasjournal}

\end{document}